\documentclass[cits]{PoS}
\pdfoutput=0
\title{Gauge fixing using overrelaxation and simulated annealing on GPUs}

\ShortTitle{Gauge fixing using overrelaxation and simulated annealing on GPUs}

\author{\speaker{Mario Schr\"ock}\\
        Institut f\"ur Physik, FB Theoretische Physik, Universit\"at
Graz, A--8010 Graz, Austria\\
       E-mail: \email{mario.schroeck@uni-graz.at}}

\author{Hannes Vogt\\
        Institut f\"ur Theoretische Physik, 72076 T\"ubingen, Germany\\
       E-mail: \email{hannes.vogt@uni-tuebingen.de}}

\abstract{We adopt CUDA-capable Graphic Processing Units (GPUs) for Coulomb, Landau and maximally Abelian gauge fixing in 3+1 dimensional SU(3) lattice gauge field theories. The local overrelaxation algorithm is perfectly suited for highly parallel architectures. Simulated annealing preconditioning strongly increases the probability to reach the global maximum of the gauge functional. We give performance results for single and double precision. To obtain our maximum performance of ~300 GFlops on NVIDIA's GTX 580 a very fine grained degree of parallelism is required due to the register limits of NVIDIA's Fermi GPUs: we use eight threads per lattice site, i.e., one thread per SU(3) matrix that is involved in the computation of a site update.
}

\FullConference{The 30th International Symposium on Lattice Field Theory\\
		 June 24--29,  2012\\
		 Cairns, Australia}

\usepackage[tight]{units}
\usepackage{amsmath}
\usepackage{wrapfig}

\newcommand{\fig}[1]{Fig.~{\ref{#1}}}
\newcommand{\tab}[1]{Tab.~{\ref{#1}}}

\newcommand{\be}{\begin{equation}}
\newcommand{\ee}{\end{equation}}

\renewcommand{\Re}{\mathfrak{Re}\,}
\newcommand{\tr}[1]{\,\text{tr}\left[#1\right]}
\newcommand{\bigO}{\mathcal{O}}
\newcommand{\e}{\text{\,e}}

\usepackage{algorithmic,algorithm}

\begin{document}

\section{Motivation}

Lattice QCD is the discrete version of the gauge theory
of the strong interaction and by construction it is invariant under local gauge
transformations of the form
\begin{equation}
	g(x) U_\mu(x) g(x+\hat\mu)^\dagger.
\end{equation}
The lattice gauge fields or \emph{link variables} $U_\mu(x)$ are connected to the continuum gauge fields
via 
\begin{equation}
	U_\mu(x) = \e^{iagA_\mu(x)}
\end{equation}
and thus live in the Lie group $\mathrm{SU}(3)$ itself instead of in its algebra.
Whereas physical observables are gauge independent,
the study of gauge dependent quantities like the fundamental
QCD Green's functions
requires to fix the gauge, i.e., to choose a specific transformation
$g(x)\in\mathrm{SU}(3)$ for all $x$.

Gauge fixing on the lattice corresponds to an optimization problem
with $\bigO(VN_c^2)$ degrees of freedom where $V=N_s^3\times N_t$ is the lattice volume.
Such being the case, the process of fixing the gauge on the lattice demands
a major part of the whole simulation's computer time and the possible
acceleration by highly parallel hardware architectures like
graphics processing units (GPUs) is clearly beneficial.
A first attempt of porting lattice gauge fixing with the overrelaxation 
algorithm to the GPU has been reported in \cite{Schrock:2011hq}.
The relaxation algorithm is particularly well suited to be accelerated
by the use of GPUs due to its strict locality which also opens the door to 
an efficient future multi-GPU parallelization. An alternative approach based on the 
steepest descent method with Fourier
acceleration has been presented in \cite{Cardoso:2012pv}.

Here we present a code package for lattice gauge fixing based on the family
of relaxation algorithms. The code is written in CUDA C++ and makes heavy use 
of template classes in order to facilitate the extension to other algorithms and 
applications.
Besides the standard relaxation algorithm \cite{Mandula:1987rh} 
our program supports overrelaxation \cite{Mandula:1990vs}
and stochastic relaxation \cite{deForcrand:1989im} to overcome the problem of critical slowing down.
Furthermore, we implemented the simulated annealing algorithm
which can be applied as a ``preconditioner'' to the gauge fields 
in order to increases the probability to reach
the global maximum of the gauge fixing functional \cite{Bali:1996dm}.

The code can be used to fix gauge configurations to the covariant
Landau gauge $\partial_{\mu} A_\mu=0$, the Coulomb gauge  $\partial_i A_i=0$
and the maximally Abelian gauge.

In the remainder of this presentation we focus on the overrelaxation algorithm
and the Landau gauge as a specific example.

\section{Algorithm}

On the lattice, gauge fixing is 
equivalent to maximizing the corresponding gauge functional, in the
case of Landau gauge
\begin{equation}
	F_g[U] = \Re\sum_{\mu, x} \tr{U^g_\mu(x)},
\end{equation}
with respect to gauge transformations $g(x)\in\mathrm{SU}(3)$ where
\begin{equation}
 U^g_\mu(x) \equiv g(x) U_\mu(x) g(x+\hat\mu)^\dagger.
\end{equation}

The relaxation algorithm optimizes the value of $F_g[U]$ locally, i.e., for all $x$ 
the maximum of $\Re\tr{g(x)K(x)}$
with
\begin{equation}\label{Kx}
	K(x):= \sum_\mu\Big( U_\mu(x) g(x+\hat\mu)^\dagger 
		+ U_\mu(x-\hat\mu)^\dagger g(x-\hat\mu)^\dagger\Big)
\end{equation}
is sought.
The local solution thereof is given by
\begin{equation}\label{localsolution}  
g(x) = K(x)^\dagger/\sqrt{\det{K(x)^\dagger}}
\end{equation}
in the case of the gauge group $\mathrm{SU}(2)$
and for $\mathrm{SU}(3)$ one iteratively operates in the three $\mathrm{SU}(2)$ subgroups
\cite{CabibboMarinari1982}.

In order to reduce the \emph{critical slowing down} of the relaxation algorithm on large
lattices, the authors of \cite{Mandula:1990vs} suggested to apply an overrelaxation algorithm which replaces
the gauge transformation $g(x)$ by $g^\omega(x),\;\omega\in[1,2)$ 
in each step of the iteration.
In practice the exponentiation of the gauge transformation is done to first or second order.

Due to the strict locality of the overrelaxation algorithm (only nearest neighbor interactions) 
we can perform a checkerboard decomposition of the lattice and 
operate on all sites of one of the two sublattices (``RED'' and ``BLACK'') at the same time.

A measure of the quality of the gauge fixing 
is the average $L_2$-norm of the gauge fixing violation $\Delta^g \neq 0$ 
\begin{equation}
\theta\equiv \frac{1}{VN_c}\sum_{x}\tr{\Delta^g(x)
\Delta^g(x)^\dagger}\,,
\end{equation}
where the sum runs over all sites $x$ and $V$ is the 
number of lattice sites.

\begin{algorithm}
	\caption{Overrelaxation}
	\label{alg1}
	\begin{algorithmic}                 
		\WHILE{precision $\theta$ not reached}
			\FOR{sublattice = RED, BLACK} 
				\FORALL{x of sublattice}
					\FORALL{SU(2) subgroups}
						\STATE $g(x) \to \sum_\mu \left\{ U_\mu^\dagger(x) + U_\mu(x-\hat\mu) \right\}$ 
							\hfill {\small $\rightarrow$ 60 Flop}
						\STATE $g(x) \to g^{\omega}(x)$, project to SU(2)
							\hfill {\small $\rightarrow$ 19 Flop}
						\FORALL{ $\mu$}
							\STATE $U_\mu(x) \to g^{\omega}(x) U_\mu(x) $
								\hfill {\small $\rightarrow$ 84 Flop}
							\STATE $U_\mu(x-\hat\mu) \to U_\mu(x-\hat\mu) g^{\omega}(x)^\dagger $
								\hfill {\small $\rightarrow$ 84 Flop}
						\ENDFOR
					\ENDFOR
				\ENDFOR
			\ENDFOR
		\ENDWHILE
	\end{algorithmic}
\end{algorithm}

The algorithm is summarized in Alg. \ref{alg1}.
In total the overrelaxation algorithm requires 751 flop per site and 
SU(2) subgroup iteration and thus 2253 flop/site for SU(3).

\section{Hardware}

We use NVIDIA's GeForce GTX 580 for our study.
The GTX 580 is the high end graphical processing unit of the Fermi architecture
which is NVIDIA's third generation of GPUs devoted to high performance
computing using NVIDIA's CUDA (Compute Unified Device Architecture)
programming environment.
Recently the successor of the Fermi architecture has been released (Kepler).
All hardware details are summarized in \tab{tab:gtx580}.

\begin{table}
	\center
	\begin{tabular}{l|c}
		architecture                 &        Fermi \\\hline
		compute capability           &       2.0\\\hline
		\# SMs (streaming multiprocessors)                        &        16\\\hline
		\# total CUDA cores           &        512\\\hline
		device memory                &         1.5 GB \\\hline
		memory bandwidth              &           192.4 GB/s\\\hline
		ECC available                &            no\\\hline
		L2 cache                     &           768 KB\\\hline
		L1 cache / SM                &          16 KB or 48 KB\\\hline
		shared memory / SM           &           16 KB or 48 KB\\\hline
		32-bit registers / SM        &           32768\\\hline
		max. registers / thread      &           63\\
	\end{tabular}
	\caption{Hardware details of the NVIDIA GeForce GTX 580.}
	\label{tab:gtx580}
\end{table}

\section{Overrelaxation on the GPU}
\subsection{First attempt}

CUDA supports natively only lattices up to three dimensions, for that reason 
we linearize the 4D lattice index
using divisions and modulo conversions of $V$ by the spatial and temporal extent
of the lattice.
We assign each lattice site to a separate thread and start 32 threads per multiprocessor.

A function which is called from the host system and which performs calculations on the GPU is called
a kernel. We implemented two kernels, one which checks the current value of the gauge fixing
functional $F_g[U]$ and the gauge precision $\theta$ after every 100th iteration step and a second 
which does the actual work,
i.e., which performs an overrelaxation step. The latter is invoked for the RED
and BLACK sublattices consecutively.

The GPU can read data from global device memory in a fast way only if the data is accurately 
coalesced: the largest memory throughput is achieved when consecutive threads
read from consecutive memory addresses. 
In order to do so we rearrange the gauge field into two blocks for the RED and BLACK
sublattices. Moreover, for the same sake of memory coalescing, we choose the site index running 
fastest which results in a storage layout in which the gauge matrices do not lie anymore in 
consecutive memory blocks.

The overrelaxation algorithm on the GPU is bandwidth bound.
Thus, in order to reduce memory traffic, we use the unitarity of $\mathrm{SU}(3)$ matrices
to reconstruct the third line of each matrix on the fly instead of reading it from global memory.
A minimal 8 parameter reconstruction \cite{Clark:2009wm} turned out to be numerically not stable
enough for our purpose since we not only have to read the gauge fields but also store them
at the end of each iteration step.

For more details about these standard tricks of GPU programming in lattice QCD 
we refer to \cite{Clark:2009wm} and references therein.

\begin{figure}[htb]
	\center
	\includegraphics[width=0.49\textwidth]{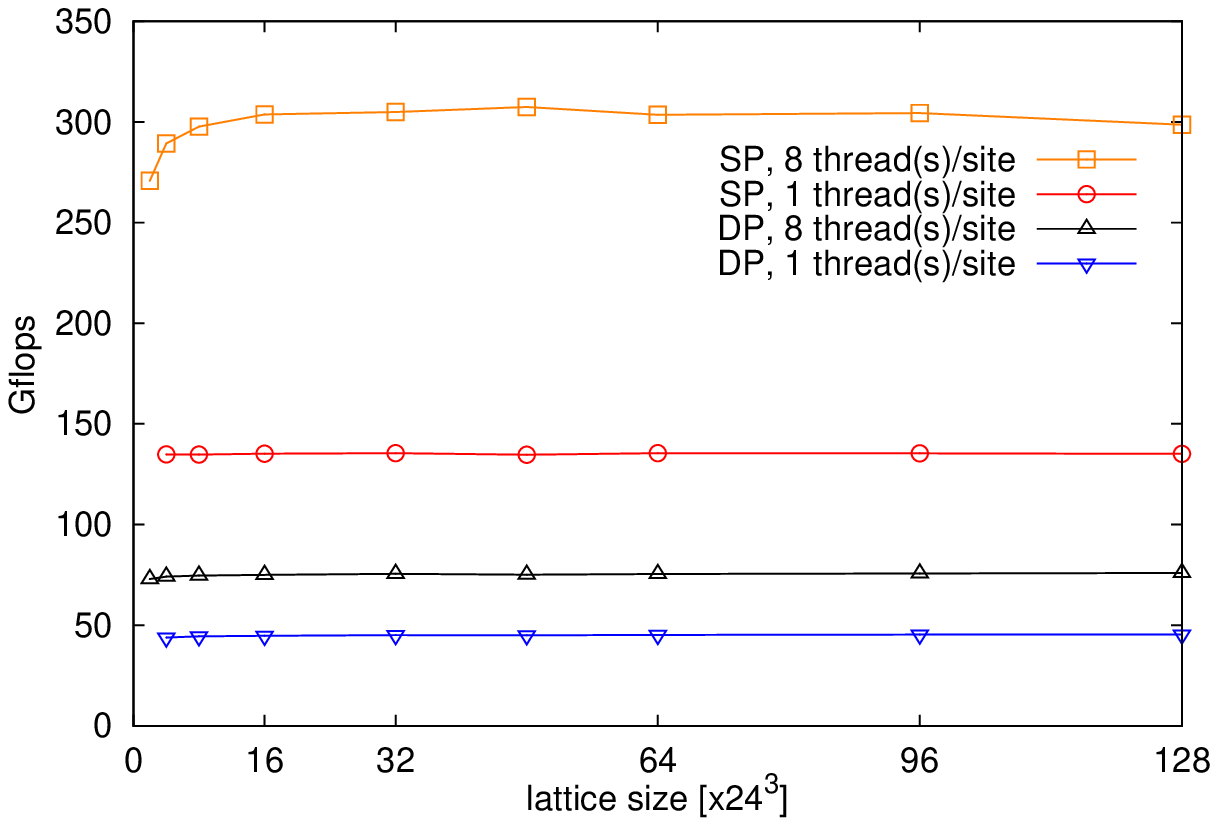}
	\includegraphics[width=0.49\textwidth]{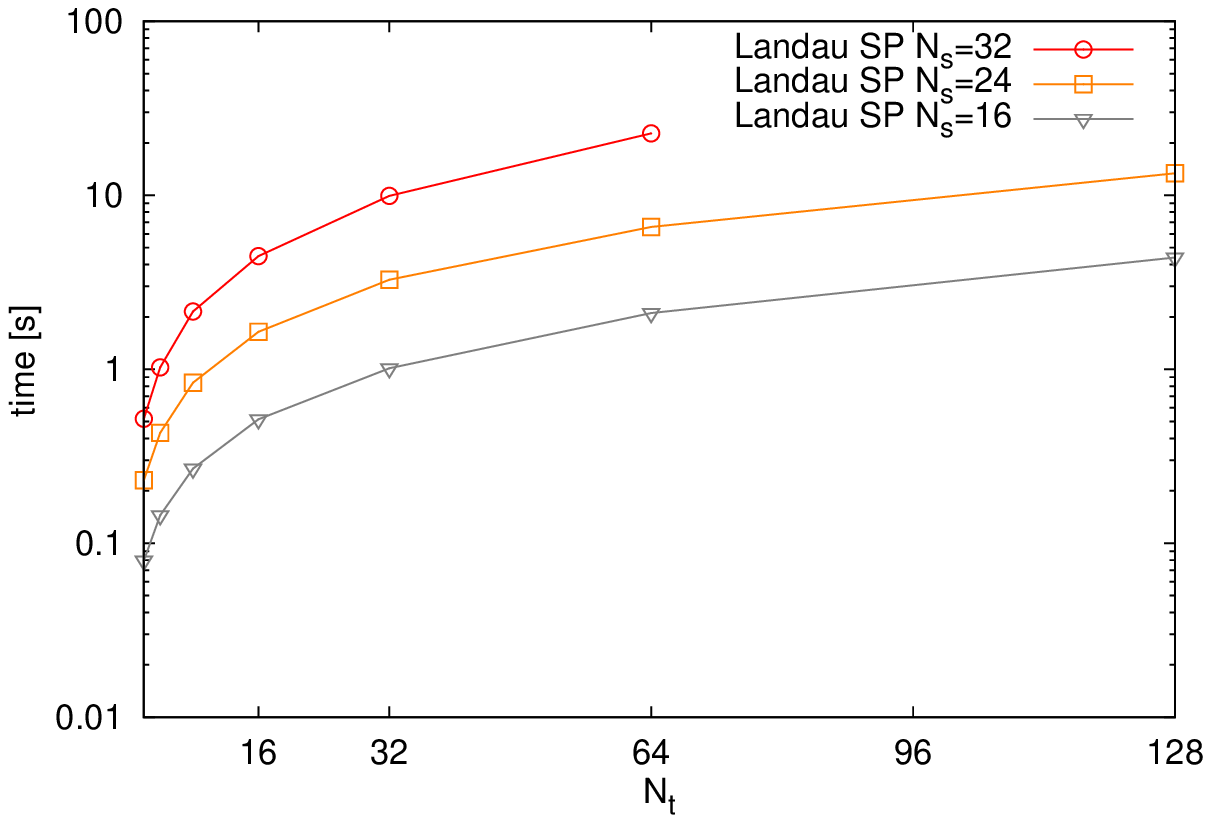}
	\caption{Left: performance for 1000 iterations. 
		Right: time needed for 1000 iterations on $N_s^3\times N_t$ lattices in SP.}
	\label{fig:performance}
\end{figure}

The performance of this first attempt can be read off from the left plot of \fig{fig:performance}.
There we show the number of flops achieved for 1000 iterations of the overrelaxation kernel
in single precision (red line, circles) and double precision (blue line, down pointing
triangles) for lattices of spatial volume $24^3$ and varying temporal extend.

\subsection{Improvement}
In the beginning of each iteration of the overrelaxation algorithm
each thread has to read its eight neighbor links from global memory 
and at the end of each iteration they have to be written
back into global memory.
These eight $\mathrm{SU}(3)$ matrices per site equal
$8\times 18$ reals = 144 reals and therewith exceed the register limit of
63 per thread (see \tab{tab:gtx580}) what results in register spills
to global memory and as a consequence negatively effect the bandwidth bound performance
of the kernel.

In order to reduce register spills we switch to a finer parallelization granularity:
instead of assigning one thread to one lattice site we now tie eight threads to 
a single lattice site, i.e., one thread for each of the eight neighbors of a site. 
Then each thread needs only 18 registers to store the gauge link.

In order to avoid warp divergences the kernel is invoked with a thread block
size of $8\times 32 = 256$. 
By doing so, each of the eight warps (warp size is 32 on the Fermi) takes care of one neighbor type
of the 32 sites and thus all threads within one warp follow the same instruction path.

The gauge transformation is then accumulated in shared memory. Since one operates on
the $\mathrm{SU}(2)$ subgroups of $\mathrm{SU}(3)$ and an $\mathrm{SU}(2)$ matrix can
conveniently be represented by four reals, this requires $4\times 32=128$ reals or 512 bytes per
thread block.

\section{Performance}
On the left hand side of \fig{fig:performance} we show that with the fine parallelization
granularity of eight threads per lattice site we achieve a maximum performance of 300 Gflops
for single precision (SP) and thus an improvement by a factor more than two compared to the
conventional one thread per site strategy.
On the right hand side of \fig{fig:performance} the time required to run 1000 iterations
of the overrelaxation algorithm on different lattice sizes is presented.

\begin{figure}[htb]
	\center
	\includegraphics[width=0.49\textwidth]{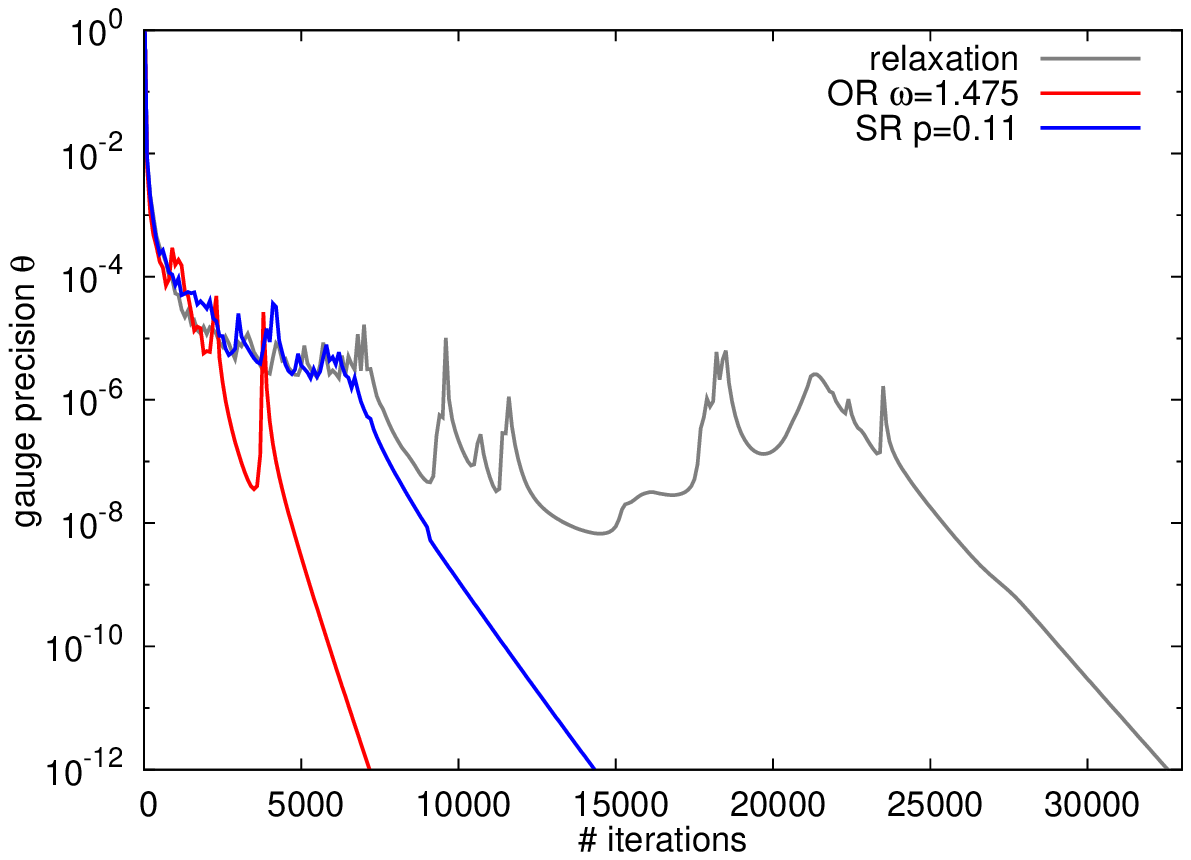}
	\includegraphics[width=0.49\textwidth]{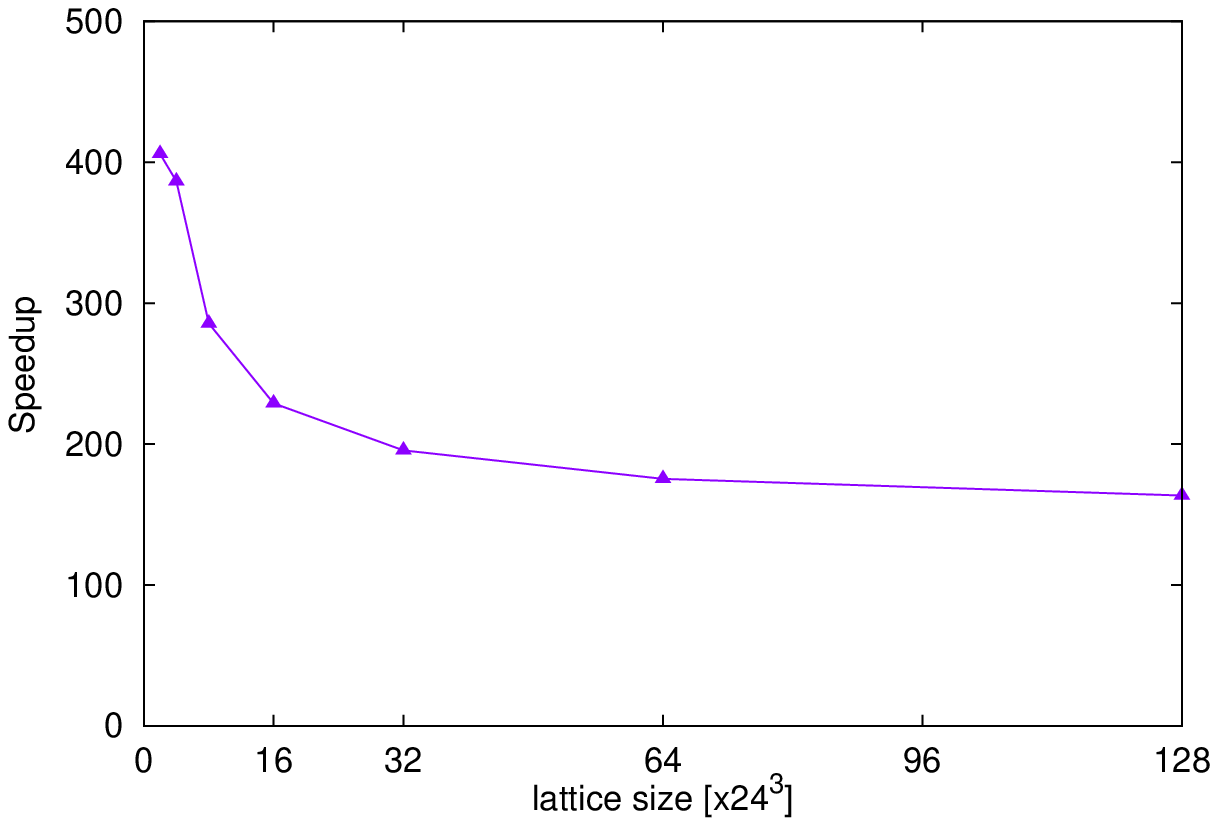}
	\caption{Left: number of iterations on a randomly chosen $\beta=6.1$, $32^4$ lattice.
	Right: speedup over CPU.}
	\label{fig:performance2}
\end{figure}

In \fig{fig:performance2}, on the left, we give an example of how tuning of the 
overrelaxation (OR) and stochastic relaxation (SR) parameters can reduce the required 
number of iterations to achieve the aimed gauge quality (here $\theta<10^{-12}$). 
This information combined with the information from \fig{fig:performance} tells
that the required time to fix the randomly chosen gauge configuration  of lattice size $32^4$ 
to the Landau gauge
with the overrelaxation algorithm
\fig{fig:performance2} to the precision of $\theta<10^{-12}$ is of the order of one minute.

Lastly, we compare our performance to the overrelaxation kernel of the
FermiQCD library \cite{FermiQCD} run in parallel with MPI on all four cores of the 
Intel Core i7-950 (``Bloomfield'') processor @ 3.07GHz.
The ratio of the time needed by FermiQCD to the time needed by our CUDA kernel
on the GTX 580 for varying lattice sizes is plotted in \fig{fig:performance2} (r.h.s.).
We find a speedup of evidently more than 150 for all lattice sizes, or in other words,
assuming linear weak scaling, the performance of our code on one GTX 580 GPU
is equivalent to the performance of the FermiQCD library on 150 Intel Core i7-950
CPUs (i.e. 600 cores) for the same algorithm.

\section{Summary}
We presented a CUDA implementation for gauge fixing on the lattice
based on the relaxation algorithms. 
In particular, our code can be used to fix gauge field configurations
to Landau, Coulomb or the maximally Abelian gauges using
simulated annealing, overrelaxation or stochastic relaxation.
Using a fine parallelization granularity of eight CUDA threads per lattice
site we achieve a maximum performance of 300 Gflops in single precision
on NVIDIA's GTX 580.
Comparing this to the performance of the overrelaxation algorithm as implemented
in the FermiQCD library run on the Intel Core i7-950 (``Bloomfield'') 
quadcore processor @ 3.07GHz in parallel using MPI, we find a speedup of more than 150.
Our code will be available for download shortly.

\begin{acknowledgments}
We thank Giuseppe Burgio and Markus Quandt for helpful discussions.
M.S. is supported by the Research Executive 
Agency (REA) of the European Union under Grant Agreement 
PITN-GA-2009-238353 (ITN STRONGnet).

\end{acknowledgments}

\providecommand{\href}[2]{#2}\begingroup\raggedright\endgroup

\end{document}